\documentclass[aip,reprint]{revtex4-1}

\usepackage{graphicx}
\usepackage{dcolumn}
\usepackage{bm}
\usepackage{hyperref}

\draft 

\begin{document}


\title{Growth and characterization of high-quality single-crystalline SnTe retaining cubic symmetry down to the lowest temperature studied} 



\author{Ayanesh Maiti}
    \affiliation{Undergraduate Department, Indian Institute of Science, Bangalore, Karnataka, India - 560012}
    \affiliation{Department of Condensed Matter Physics and Materials Science, Tata Institute of Fundamental Research, Mumbai, Maharashtra, India - 400005}
\author{Ankita Singh}
\author{Kartik K. Iyer}
\author{Arumugam Thamizhavel}
\affiliation{Department of Condensed Matter Physics and Materials Science, Tata Institute of Fundamental Research, Mumbai, Maharashtra, India - 400005}


\date{\today}

\begin{abstract}
SnTe, an archetypical topological crystalline insulator, often shows a transition from a highly symmetric cubic phase to a rhombohedral structure at low temperatures. In order to achieve the cubic phase at low temperatures, we have grown SnTe employing the modified Bridgman method and studied its properties in detail. Analysis of the crystal structure using Laue diffraction and rocking curve measurements show high degree of single crystallinity and mosaicity of the sample. The magnetic susceptibility shows diamagnetic behaviour and the specific heat data matches the phonon contributions typical of a bulk insulator. Resistivity data exhibit metallic conduction similar to two-dimensional systems and the signature of structural transition has not been observed down to the lowest temperature studied. Detailed powder $x$-ray diffraction measurements show cubic structure in the entire temperature range studied. This is supported by the ARPES data at low temperature exhibiting a Dirac cone typical of a topological material. These results demonstrate that the ground state structure of SnTe can be stabilized in the cubic phase, providing a promising platform for quantum applications. 
\end{abstract}


\maketitle 

\begin{figure}

\includegraphics[width=0.49\textwidth]{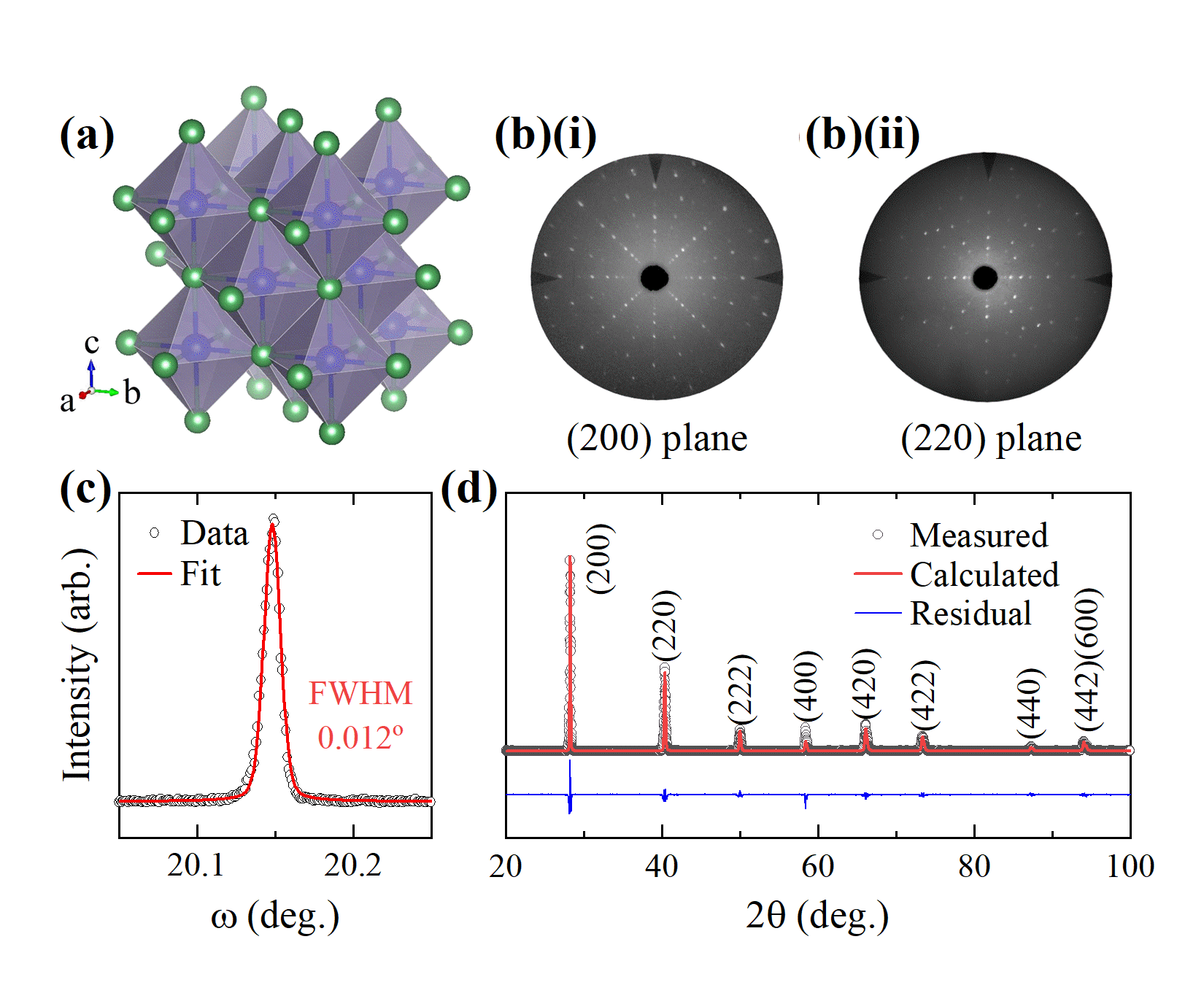}
\caption{\label{Fig1}(a) Crystal structure of SnTe. Sn (green balls) and Te atoms (blue balls) form in the cubic rock salt structure. (b) Laue diffraction patterns in (i) $\langle200\rangle$ and (ii) $\langle220\rangle$ sample orientations. (c) Rocking curve of the $(220)$ reflection (open circles) and its fit with a Voigt function (line). (d) Powder XRD pattern collected at room temperature (symbols). Rietveld fit of the data is shown by line superimposed over the experimental data. The residue is shown below the plot.}
\end{figure}

Topological insulators are a class of bulk-insulators that contain symmetry-protected metallic states at the surface \cite{TM, Ando}. These states behave like Dirac Fermions and have linear band dispersions. Strong spin-orbit coupling (SOC) breaks the spin-degeneracy of these surface bands, forcing surface currents to be spin-polarized. This effect finds many applications in the development of next-generation spintronics and quantum computers \cite{TI-app}. Some materials behave like topological insulators only in their crystalline forms \cite{TCI}. Tin telluride (SnTe) is one such {\it topological crystalline insulator} \cite{SnTe-TCI1,SnTe-TCI2,ARPES} that has attracted a lot of interest. SnTe crystallizes in NaCl structure with the space group $Fm\bar{3}m$ at room temperature. In this structure, the tin atoms occupy FCC lattice sites and tellurium atoms occupying their octahedral voids, as shown in Fig.~\ref{Fig1}(a). Previous studies have shown structural transitions to a relatively lower symmetric structure at low temperatures or high pressures \cite{Pres-st,SnTe-res1,Rhomb1,Rhomb2}. Under atmospheric pressure, SnTe exhibits a rhombohedral distortion at low temperatures where the transition temperature is reported to depend on the charge carrier concentration \cite{SnTe-res1,Rhomb1,Rhomb2}. The bulk properties of SnTe is insulating and have also been extensively studied for applications in efficient, eco-friendly thermoelectrics \cite{SnTe-TE}.

The topological behaviour in SnTe arises from its non-magnetic nature and NaCl-like cubic crystal structure consisting of several mirror planes that satisfy various symmetry protections. Some of the mirror symmetries will be broken due to the transition to a less symmetric rhombohedral structure. Therefore, the topological states will not survive on some of the surfaces at low temperatures. This has significant implications in its electrical transport behaviour. In particular, the structural transition could be tracked by a kink in the transport data \cite{SnTe-res1,SnTe-res2}. So far, the specific heat is studied only in the low-temperature regime (below $60$~K) \cite{SnTe-HC}. The defects in the pristine compound (loss of Sn during sample preparation) and chemical disorder makes the system more complex. In this study, we succeeded to grow high-quality single crystals of SnTe with a reduced degree of disorder and defects. We investigated the crystal structure, magnetic susceptibility, specific heat and electrical transport in the temperature range 2-300~K. Our results show typical insulating bulk behaviour and the survival of the cubic symmetry down to the lowest temperature studied. 

SnTe single crystal was grown using the modified Bridgman growth technique. A precisely weighed stoichiometric mixture of $99.999$\% pure elemental tin and tellurium was taken in a point-bottom quartz tube and sealed in a high vacuum of about 10$^{-6}$~mbar. The sealed ampoule was placed in a box-type resistive heating furnace and heated to 1050~$^{\circ}$C at 50~$^{\circ}$C/hr, where it was homogenized for 24 hrs. Then, it was fast-cooled at a rate of 100~$^{\circ}$C/hr down to 850~$^{\circ}$C. This point was chosen well above the melting temperature of SnTe to avoid premature solidification due to thermal inhomogeneities arising from fast cooling. After a brief rest, the furnace was cooled down to 650~$^{\circ}$C, well below the melting point of SnTe, at a very slow cooling rate of 1~$^{\circ}$C/hr. The point bottom quartz ampoule helped in the nucleation of the crystal, which further grew into a large boule during the slow cooling process. Finally, the sample was cooled down to room temperature at a moderate cooling rate of 30~$^{\circ}$C/hr in order to avoid thermal shock to the grown crystal.

The phase purity of our single crystals was confirmed through powder $x$-ray diffraction (XRD) measurements conducted in Bragg-Brentano geometry with a filtered Cu-$K\alpha$ $x$-ray source using a PANalytical $x$-ray diffractometer. Further powder XRD measurements were carried out at selected temperatures in the range 4-300~K using a Rigaku SmartLab machine equipped with a low-temperature attachment. The crystal structure was evaluated through a Rietveld refinement \cite{Rietveld} of the data, carried out using the FULLPROF software suite. Laue diffraction measurements were performed in the back-reflection geometry to examine the single crystalline nature and orient the crystal along its $\langle200\rangle$ and $\langle220\rangle$ crystallographic directions. The aligned crystal was cut into suitable pieces for low-temperature magneto-transport measurements using a spark erosion cutting machine. The crystal quality was examined through rocking curves measured in a Rigaku SmartLab machine. The temperature dependence of various physical properties such as heat capacity, electrical resistivity, magnetic susceptibility, etc., were measured using PPMS, MPMS and SQUID magnetometer devices from Quantum Design. The reproducibility of the data has been confirmed through measurements on different sample pieces.

The Laue diffraction patterns corresponding to the $\langle200\rangle$ and $\langle220\rangle$ orientations of the sample are shown in Fig.~\ref{Fig1}(b). We observe sharp and well-defined circular spots exhibiting the diffraction patterns for an FCC lattice. The high quality of the crystals is evident from these data. The rocking curve for the $\langle220\rangle$ reflection is shown in Fig.~\ref{Fig1}(c). The very sharp peak with full width at half maximum of about $0.012^{\circ}$ confirms the high degree of crystallinity and mosaicity. The powder XRD pattern, shown in Fig.~\ref{Fig1}(d), exhibits sharp peaks at diffraction angles ($2\theta$) matching the NaCl-like cubic structure with a lattice constant of $6.316$~\AA. There are no traces of any discernible impurity peaks suggesting a highly phase-pure sample.

\begin{figure}
\includegraphics[width=0.49\textwidth]{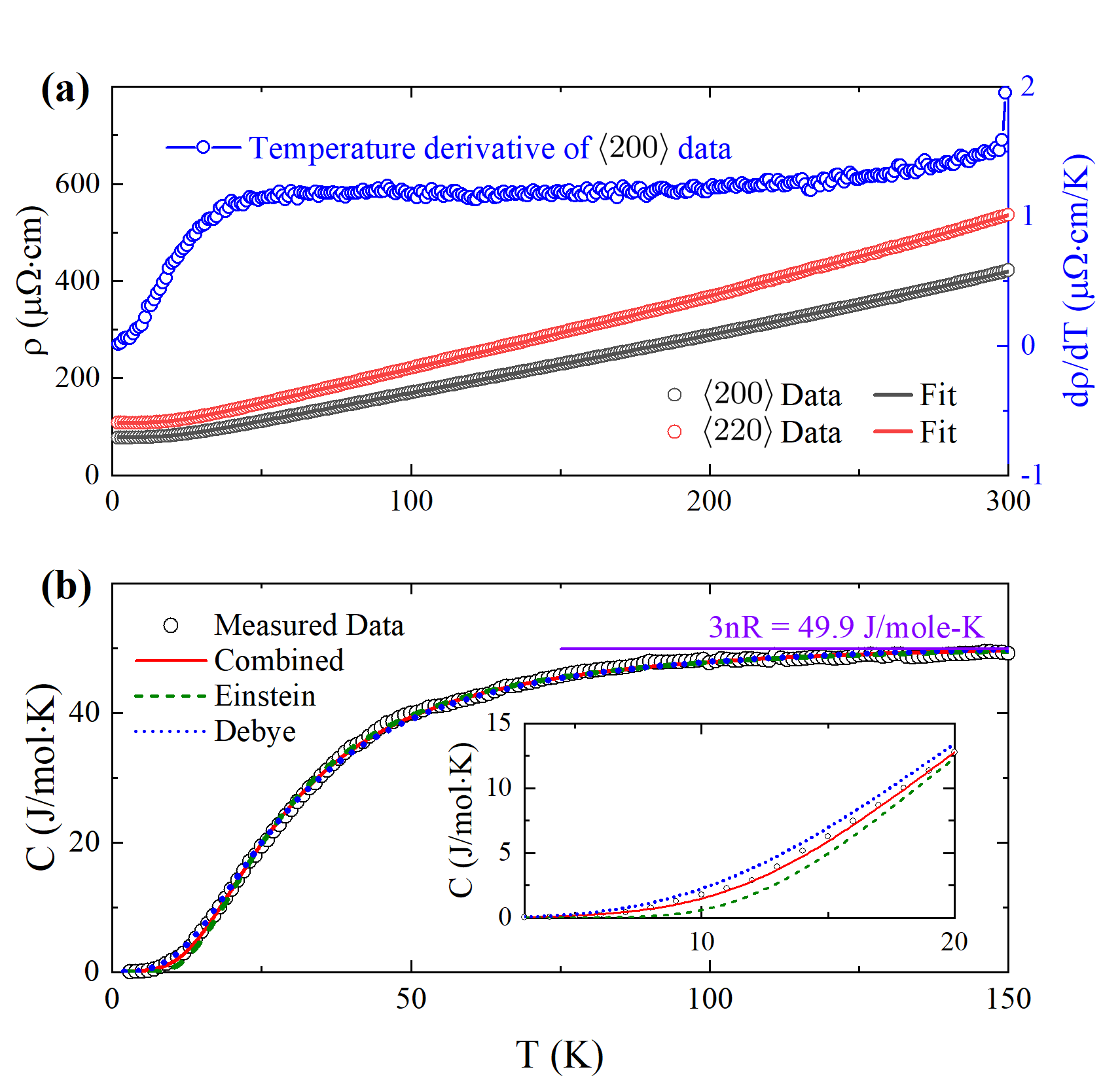}
\caption{\label{Fig2}(a) Left axis: Electrical resistivity $\rho$ data and fit along $\langle200\rangle$ and $\langle220\rangle$ directions of our sample. Right axis: Temperature derivative of resistivity along $\langle200\rangle$ direction. (b) Specific heat $C$ data as a function of temperature and its fits with the Debye, Einstein and combination models. The inset shows the low temperature fit quality.}
\end{figure}

The structural transition in SnTe, which is generally tracked through a kink in the resistivity data, usually produces a peak-like feature in its temperature derivative \cite{Rhomb2}. In Fig.~\ref{Fig2}(a), we show the resistivity as a function of temperature. The experimental data exhibit a metallic temperature dependence consistent with earlier observations \cite{SnTe-res1,SnTe-res2}. Since the material is expected to be insulating in the bulk, the metallic conduction and its temperature dependence may be attributed to the topologically ordered surface states as supported by angle-resolved photoemission (ARPES) measurements \cite{SnTe-TCI1,SnTe-TCI2,ARPES}. We model the data following the analysis used for the topological insulator, Bi$_2$Te$_3$ \cite{Bi2Te3-res}. The low temperature $T^4$ term is extended to higher temperatures using the Bloch-Gr\"{u}neisen relation for two dimensional (2D) metals \cite{BG}.
\begin{equation}
\rho(T) = \rho_{\rm d} + \rho_{\rm ep} \left(\frac{T}{\Theta}\right)^4 \int_{0}^{\Theta/T} \frac{x^4 {\rm e}^x dx}{({\rm e}^x-1)^2} + \rho_{\rm ee} T^2
\label{Eqn5}
\end{equation}
where $\Theta$ is the characteristic temperature. The first term $\rho_{\rm d}$ corresponds to the contribution from bulk conduction and contact resistance if there is any. The second term accounts for electron-phonon scattering effects described by the Bloch-Gr\"{u}neisen relation for 2D metals, with $\rho_{\rm ep}$ as a scaling constant. The last term represents the effect of Umklapp scattering processes in a low-temperature regime, with $\rho_{\rm ee}$ as a scaling factor. Fig.~\ref{Fig2} shows the resulting best fits. The model shows a good agreement with the experimental data and an estimate of the Bloch-Gr\"{u}neisen temperature, $\Theta$ is found to be $144$~K for both the orientations.

\begin{figure}
\includegraphics[width=0.49\textwidth]{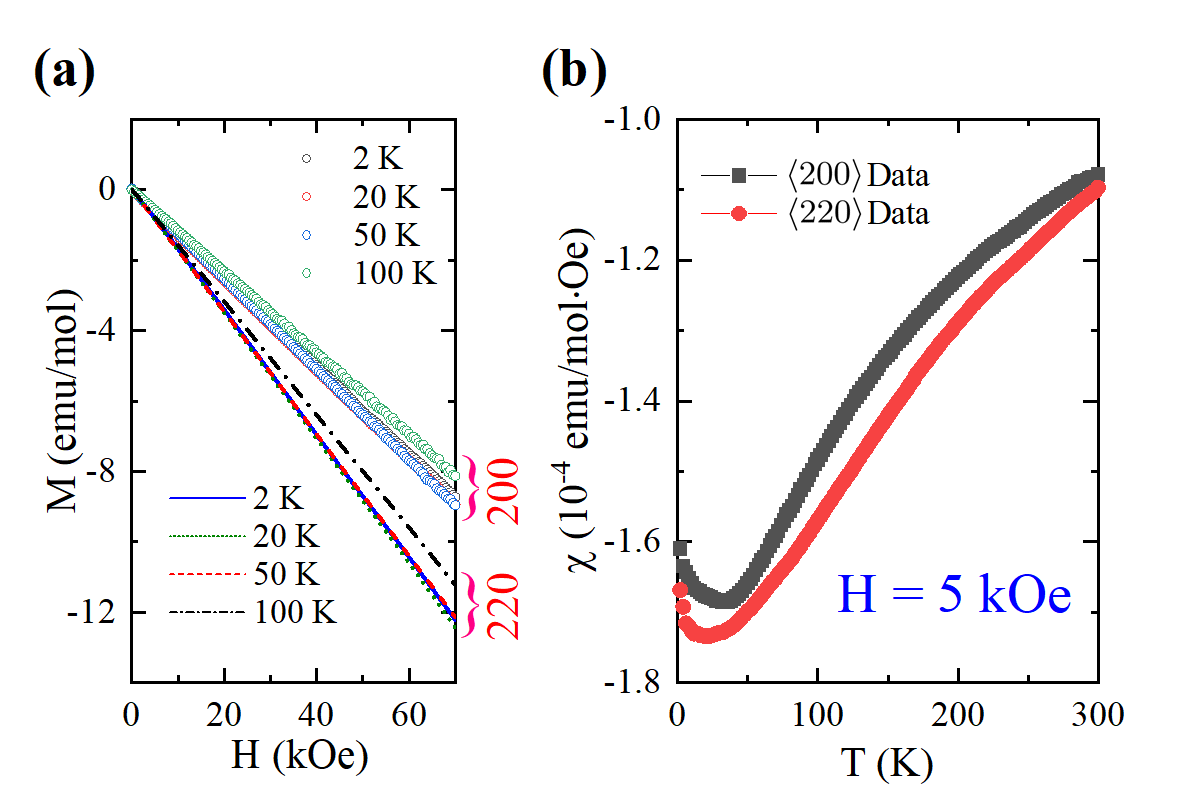}
\caption{\label{Fig3}(a) Magnetization, $M$ measured with the magnetic field, $H$ applied along $\langle200\rangle$ and $\langle220\rangle$ directions at different temperatures. (b) Magnetic susceptibility, $\chi$ as a function of temperatures measured in a magnetic field $H = 5$~kOe in $\langle200\rangle$ and $\langle220\rangle$ orientations.}
\end{figure}

The specific heat, $C$ of the sample measured as a function of temperature is shown in Fig.~\ref{Fig2}(b). $C$ increases with the increase in temperature to about $50~{\rm J/mol \cdot K}$ above $100$~K and then gradually saturates at higher temperatures. The data do not show any features in the whole temperature range studied that could be associated with a structural transition in the crystal. The high-temperature value agrees well with the prediction from the Dulong-Petit law; $3nR = 49.9$~${\rm J/mol \cdot K}$. Here $R$ is the ideal gas constant and $n$ is the atoms per formula unit ($n = 2$ for SnTe). The experimental heat capacity data can be simulated reasonably well within the Debye and Einstein models of phonon specific heat described in Eqns.~(\ref{Eqn2})~and (\ref{Eqn3}) respectively:
\begin{equation}
C_{\rm Deb} (T,\Theta_{\rm D}) = 9nR\left(\frac{T}{\Theta_{\rm D}}\right)^3 \int_0^{\Theta_{\rm D}/T}\frac{x^4{\rm e}^xdx}{({\rm e}^x-1)^2}
\label{Eqn2}
\end{equation}
\begin{equation}
C_{\rm Ein} (T,\Theta_{\rm E}) = 3nR\left(\frac{\Theta_{\rm E}}{T}\right)^2 \frac{{\rm e}^{\Theta_{\rm E}/T}}{({\rm e}^{\Theta_{\rm E}/T}-1)^2}
\label{Eqn3}
\end{equation}
where $\Theta_{\rm D}$ and $\Theta_{\rm E}$ are the respective Debye and Einstein temperatures. The estimated values of $\Theta_D$ and $\Theta_E$ are $120$~K and $90$~K respectively. The simulated values do not match the experimental data well at low temperatures. The results within the Debye model is somewhat larger than the experimental values and the data from the Einstein model goes below the experimental values as shown in the inset of Fig. \ref{Fig2}(b). To achieve a better description, we fit the specific heat with a linear combination of the two models:
\begin{equation}
C_{\rm Com}(T) = \alpha C_{\rm Deb} (T,\Theta_{\rm D}) + (1-\alpha)C_{\rm Ein} (T,\Theta_{\rm E})
\end{equation}
This model provides a good description of the experimental data for $\Theta_{\rm D} = \Theta_{\rm E} = 100$~K and $\alpha = 35.4\%$. The Debye model captures the low energy phonon contributions, whereas the Einstein model predicts the behaviour when all phonon modes are active. The analysis in the present case suggests that both the contributions are significant at moderately low temperatures, which is also reflected by the low characteristic Debye/Einstein temperature. Evidently, the heat capacity of this material could be captured well using the phonon contributions as expected from an insulating bulk sample. 

\begin{figure}
\includegraphics[width=0.49\textwidth]{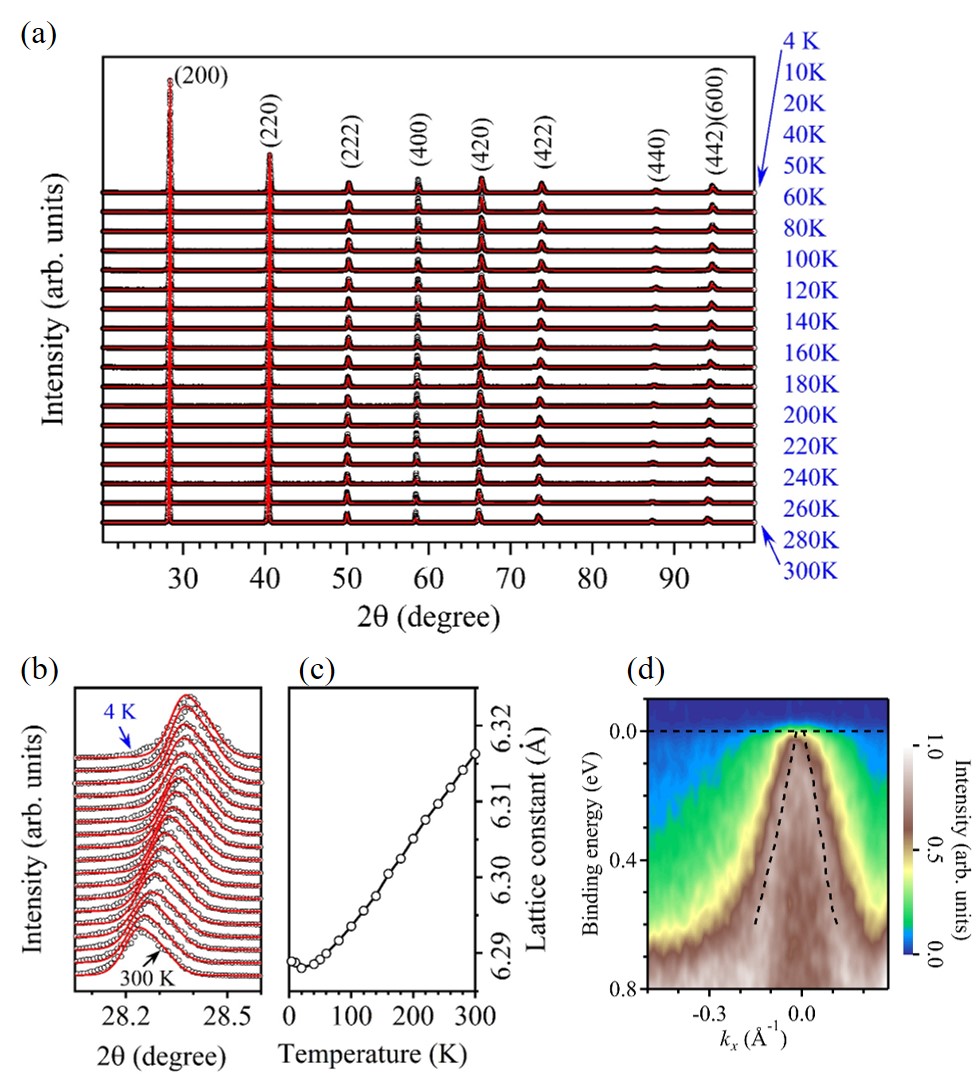}
\caption{\label{Fig4}(a) Powder XRD patterns of our sample (circles) measured at selected temperatures along with their Rietveld fits (lines). (b) (200) Bragg peak in expanded angle scale to demonstrate the thermal effects. (c) Thermal evolution of the lattice constant. (d) ARPES data collected at 22 K showing a Dirac cone due to the surface states. Dashed lines represent the peak positions in each branch of the Dirac cone.}
\end{figure}

The isothermal magnetization, $M$ as a function of the applied magnetic field, $H$ measured at different temperatures is shown in Fig.~\ref{Fig3}(a). The behaviour is linear in the entire field range (0-70 kOe) with a small anisotropy implying that the magnetic susceptibility is independent of the applied field. The magnetization as a function of temperature at a field of 5~kOe is shown in Fig.~\ref{Fig3}(b). The negative susceptibility observed in Fig. \ref{Fig3} indicates diamagnetic behaviour arising essentially from the closed-shell electronic configurations (Larmor diamagnetism). There may be some contributions from the conduction electrons (Landau diamagnetism). The susceptibility remains diamagnetic in the entire temperature range studied. The magnitude of susceptibility gradually enhances on cooling and exhibits an upturn below about $50$~K. The origin of such behaviour is not clear at present. There is some anisotropy in the magnetization that changes with the temperature.

The temperature dependence of the powder XRD pattern is shown in Fig.~\ref{Fig4}(a). We observe intense sharp peaks corresponding to the cubic structure of the sample in the whole temperature range. Usually, the lowering of symmetry leads to a splitting of diffraction patterns. We did not observe any such effects down to $4$~K studied. Moreover, the observed XRD pattern could not be derived using a rhombohedral structure. This suggests that there is no rhombohedral distortion in the temperature range 4-300 K, consistent with the resistivity and specific heat data discussed above. There is a slight decrease in the Bragg angle with raising of temperature, as demonstrated in Fig.~\ref{Fig4}(b), with similar trends in all the reflections. The Rietveld fits are shown as solid lines superimposed over the experimental data points in Fig.~\ref{Fig4}(a)-(b), matching the cubic crystal structure excellently. Clearly, the temperature variation primarily reflects thermal expansion without any change in the crystal structure type. Fig.~\ref{Fig4}(c) presents the extracted lattice constants as a function of measured temperatures. As we cool the sample from 300 K, the lattice parameter decreases almost linearly with temperature till about 50 K, reflecting a constant coefficient of thermal expansion of about $17.4\pm0.1$ ppm/K. Below 50 K, the behaviour is somewhat different. The peak shift shown in Fig.~\ref{Fig4}(b) reflects a similar trend.

In order to verify the survival of topological states at the surface at low temperatures, the energy band structure is probed by angle-resolved photoemission spectroscopy (ARPES). The experiments were done using a DA30L analyzer at 22 K on the (001) surface of the same sample cleaved in ultrahigh vacuum ($\sim$10$^{-11}$ torr).\cite{ARPES} The experimental results are shown in Fig.~\ref{Fig4}(d) exhibiting an intense Dirac cone due to the surface states within the energy gap of the bulk bands; these results are consistent with the results at higher temperatures.\cite{SnTe-TCI1,SnTe-TCI2} These results establish that the topologically ordered surface states survive down to the lowest temperatures studied so far.

In conclusion, we have studied the growth and characterization of a high-quality single crystal of SnTe. To stabilize the highly symmetric cubic structure down to low temperature and improve crystal quality, we have employed the modified Bridgman method in a specially designed quartz ampule to grow our sample. Rocking curve analysis and Laue diffraction pattern show a high degree of single crystallinity and mosaicity of the sample. The temperature dependence of the specific heat could be captured well within the combined Debye and Einstein models manifesting essentially phonon contributions to the heat capacity of this material as expected in a bulk semiconducting material. The magnetic susceptibility data show diamagnetic behaviour and corroborate well with the conclusions from the specific heat data. The electrical resistivity shows metallic temperature dependence; the temperature evolution of resistivity does not show any signature of the structural transition. The detailed analysis of the XRD patterns shows cubic rock salt structure in the entire temperature range studied down to $2$~K. All these results establish that the ground state structure of SnTe can be stabilized with its cubic symmetry, allowing access to the low-temperature regime available for quantum behaviour of this material which is essential for engineering quantum devices.


%
%

%

\begin{acknowledgments}
The authors acknowledge financial support from the Department of Atomic Energy (DAE), Government of India. A.M. thanks the Department of Science and Technology, Govt. of India for financial support under the KVPY program.
\end{acknowledgments}

\section*{Author Declarations}
The authors have no conflicts to disclose.

\section*{Data Availability}
The data that supports the findings of this study are available within the article.

\bibliography{ref.bib}

\end{document}